\documentstyle[twoside,fleqn,espcrc2,epsfig]{article}
\newcommand{\noi}{\noindent}
\newcommand{\eq}{\begin{equation}}
\newcommand{\en}{\end{equation}}
\newcommand{\eqa}{\begin{eqnarray}}
\newcommand{\ena}{\end{eqnarray}}
\def\Journal#1#2#3#4{{#1}{\bf #2} (#4) #3}

\def\PLB{Phys. Lett.~{\bf B}}
\def\PRL{Phys. Rev. Lett.~}
\def\PRD{Phys. Rev. ~{\bf D}}

\def\RMP{Rev. Mod. Phys.~}

\def\SJNP{Sov. J. Nucl. Phys.~}

\title{
$SU(2)$ lattice gauge theory at non-zero temperature with fixed
holonomy boundary condition
\thanks{Talk presented by M.~M\"uller-Preussker}
}
\author{
E.-M. Ilgenfritz%
\address{Research Center for Nuclear Physics, Osaka University,
Osaka 567-0047, Japan},
B. Martemyanov%
\address[ITEP]{Institute for Theoretical and Experimental Physics,
Moscow 117259, Russia},
M. M\"uller-Preussker%
\address{Humboldt-Universit\"at zu Berlin, Institut f\"ur Physik,
Invalidenstr. 110, D-10115, Germany}%
\thanks{Supported by EU-TMR network FMRX-CT97-0122},
A. I. Veselov%
\addressmark[ITEP]
}
\begin{document}
\begin{abstract}
We study $SU(2)$ lattice gauge theory at $T>0$ in a finite
box with fixed holonomy value at the spatial boundary.
We search for (approximate) classical solutions
of the lattice field equations and find in particular
the dissociated calorons recently discussed by van
Baal and collaborators.
\vspace{1pc}
\end{abstract}

\maketitle

The quark confinement has not yet found a satisfactory
explanation. Several models are under consideration.
The dual superconductor scenario views confinement as a dual 
Meissner effect due to the condensation of Abelian monopoles.
An alternative promising
approach is based on the center-vortex dominance picture.
On the other hand there is the semiclassical approach based on
instanton solutions. It provides successful phenomenology
for many phenomena in hadron physics.
Unfortunately, instanton gas or liquid models fail to explain
confinement. The question arises, whether other extended classical
objects - e.g. monopoles or dyons - could be suited to describe 
confinement within a semi-classical approach.

We consider $SU(2)$ lattice gauge theory at finite temperature
with periodic boundary conditions characterized additionally by a
non-trivial holonomy ${\cal P}({\bf x})$ at the spatial boundary.
The Polyakov line at the boundary is then the trace of the holonomy
$L({\bf x})=\frac{1}{2}{\mathrm tr} {\cal P}({\bf x}).$

Our investigation (see also \cite{we}) is motivated by papers of
Pierre van Baal and co-workers \cite{vb_dubna}, who have thoroughly
reconsidered the carriers of topological charge
in a Yang-Mills field at finite temperature (calorons).
They have analytically demonstrated that completely
different caloron solutions appear once a non-trivial holonomy
${\cal P}({\bf x})$ at ${|\bf x}|\rightarrow\infty$ is admitted.
These solutions differ from the 't Hooft periodic instantons
employed for the standard semi-classical approach at finite temperatures
\cite{t_hooft,harrington_shepard,gross_pisarski_yaffe}. The latter
solutions have trivial holonomy {\it i.e.}
${\cal P}({\bf x}) \rightarrow 1$ for ${|\bf x}|\rightarrow\infty$.

The most interesting feature of the {\it new calorons} is the fact that
monopole constituents of an
instanton can become explicit as degrees of freedom
\cite{instanton_constituents,caloron_lattice}. They carry magnetic
charge (in fact, they are BPS monopoles \cite{bps})
and $1/N_{\mathrm{color}}$ units of topological charge. Being
part of classical solutions of the Euclidean field equations,
one can hope that the instanton constituents can play an independent
role in the semiclassical analysis of $T\ne0$ Yang-Mills theory
(and of full QCD).

Here we present an exploratory study where we have searched
for characteristic differences between the two phases as far as
semiclassical background fields are concerned. The latter become visible
in the result of cooling.

We have fixed during the simulation and under cooling the boundary
time-like link variables in order to keep a certain value of
${\cal P}({\bf x})={\cal P}_{\infty}$ everywhere on the spatial
surface of the system while conserving periodicity.
In this case, the influence of the respective phase, that we want to
describe, is twofold : ({\it i}) the cooling starts from genuine thermal
Monte Carlo gauge field configurations, generated on a $N_s^3\times N_t$
lattice; ({\it ii}) the value of the holonomy ${\cal P}_{\infty}$ was
chosen in accordance with the average of $L$, which is approximately
vanishing in the confinement phase and nonvanishing but far from unity 
in the deconfinement phase at not too high temperatures.

For a lattice of size $16^3 \times 4$ we have chosen 
$\beta=2.2$ (confinement, $\langle L \rangle \simeq 0.$) and 
$\beta=2.4$ (deconfinement, $\langle L \rangle =0.27$),
respectively. We freeze the
timelike links $U_{x,\mu=4}$ at the spatial boundary
equally to each other such that $(U_{x,\mu=4})^{N_t}={\cal P}_{\infty}$.
For the holonomy itself, an `Abelian' form
${\cal P}_{\infty}=a_0 + i~a_3~\tau_3$ was chosen, with
$a_0=~\langle L \rangle$ and $a_3=\sqrt{1-a_0^2}$ in correspondence with
the average Polyakov line. 
In order to search exclusively for objects with low action
the criterion for stopping at some
cooling step $n$ was that $S_n < 2~S_{\mathrm{inst}}$, the last change
of action $|S_n - S_{n-1}| < 0.01~S_{\mathrm{inst}}$, and 
$S_n-2~S_{n-1}+S_{n-2} < 0$ ($S_{\mathrm{inst}}$
denoting the action of a single instanton).
For each $\beta$-value we have scanned $O(200)$ configurations obtained
by cooling.

As can be seen from Table 1 the cooled sample obtained  
in the confinement phase has a 
different composition than that of the deconfinement phase.

\begin{table}[ht]
\caption{Relative frequencies of the occurence of different kinds of
(approximate) solutions for $\beta=2.2$ (confinement) and
$\beta=2.4$ (deconfinement) after cooling.}
\vspace{5mm}
\begin{center}
\begin{tabular}{lcc}
\hline
Type of solution     & $ \beta=2.2 $        &  $ \beta=2.4 $        \\
\hline
$DD$                 & $ 0.63  \pm 0.08  $  &  $ 0.02  \pm 0.01  $  \\
$D\overline{D}$      & $ 0.27  \pm 0.05  $  &  $ 0.78  \pm 0.07  $  \\
$CAL$                & $ 0.02  \pm 0.01  $  &  $ 0.              $  \\
$M$, $2M$            & $ 0.01  \pm 0.01  $  &  $ 0.07  \pm 0.02  $  \\
trivial vacuum       & $ 0.07  \pm 0.03  $  &  $ 0.13  \pm 0.03  $  \\
\hline
\end{tabular}
\end{center}
\end{table}
\noi
In the following let us explain these configurations in some detail.

\begin{figure}
 \begin{minipage}{6.0cm}
 \begin{center}
  \epsfig{file=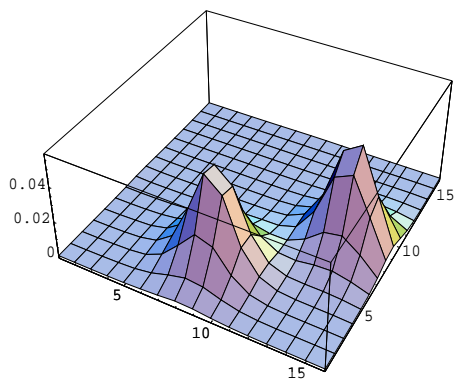,width=6.0cm,height=5.0cm} 
 \end{center}
 (a)
 \end{minipage}
\vspace{6mm}
 \begin{minipage}{6.0cm}
  \begin{center}
   \epsfig{file=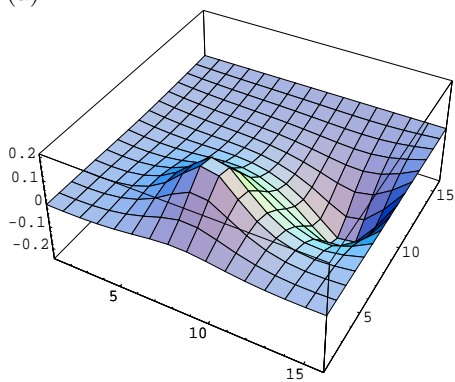,width=6.0cm,height=5.0cm}
  \end{center}
  (b)
  \end{minipage}
  \vspace{3mm}
 \caption{
The $2d$ projected distributions of topological charge (a) and
Polyakov line (b) for a selfdual $DD$ pair.}
\label{fig.1&2}
\end{figure}

\medskip\noi
In the confinement phase clearly dominate 
`dyon-dyon' pairs ($DD$) reminiscent of the {\it new caloron} solutions.  
In Figs. \ref{fig.1&2} we show, projected onto the
$x_1-x_2$-plane ({\it i.e.} summed over $x_3,x_4$ or $x_3$, resp.),
the topological charge and the Polyakov line, respectively,
of such a `dyon' pair. Notice the opposite sign of the Polyakov line 
near the maxima of the two bumps of topological charge.

Other selfdual objects, having a rather $O(4)$ rotationally invariant
distribution of action and topological charge, are frozen out relatively
infrequently. They resemble the 't Hooft periodic instanton. We call them
caloron ($CAL$). Under the specific boundary
conditions, however, the Polyakov line distribution around the caloron
exhibits opposite peaks. Thus, 
this type of configurations appears to be a limiting case of the
`dyon-dyon' pairs.

\begin{figure}
 \begin{minipage}{6.0cm}
\begin{center}
\leavevmode
\epsfxsize = 5.5cm
\epsffile[40 50 540 690]{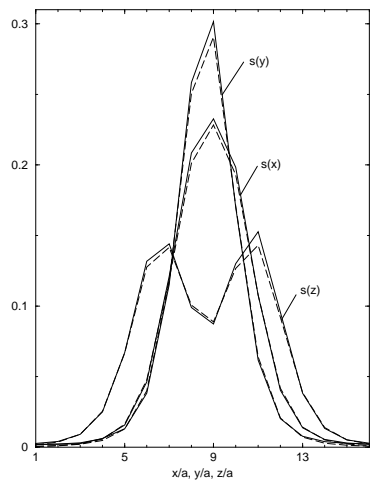}
\end{center}
 \end{minipage}
\vspace{3mm}
 \caption{
Action density profiles $~s(x), s(y), s(z)~$ of a $DD$ event.
Dashed lines correspond to a fit with van Baal's solution (scale size
$~\rho \cdot T = 0.63~$).}
\label{fig.3}
\end{figure}

Both $DD$ and $CAL$ objects can be well fitted by the analytically 
known solutions \cite{caloron_lattice}. 
In Fig. \ref{fig.3} we show measured and fitted action density profiles 
for a typical $DD$ configuration found in the confinement phase.

\begin{figure}
 \begin{minipage}{6.0cm}
\begin{center}
\leavevmode
\epsfxsize = 5.5cm
\epsffile[40 50 540 690]{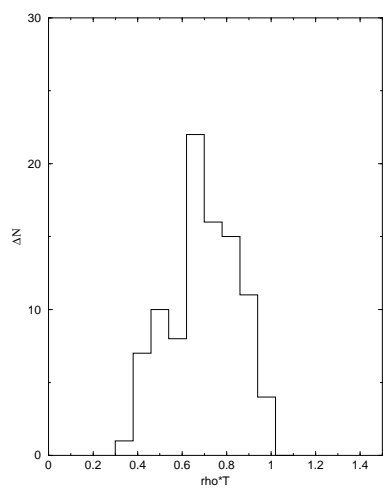}
\end{center}
 \end{minipage}
\vspace{3mm}
 \caption{
$\rho$ distribution of $94 ~~DD$ and $CAL$ events, resp.
}
\label{fig.4}
\end{figure}

The frequency distribution of the scale sizes $\rho$ 
obtained by these fits is shown in Fig. \ref{fig.4}. 

Mixed configurations with two lumps of opposite topological charge
are found in a quarter of the configurations. We call them `dyon-antidyon'
pairs ($D\overline{D}$). 
In these configurations the Polyakov line has a same-sign maximum on
top of the opposite-sign topological charge lumps. Besides of this,
the two sums $Q_{+}=\sum_x q(x)~\Theta(q(x))$ and
$Q_{-}=\sum_x q(x)~\Theta(-q(x))$ are almost equal to $+\frac12$ and
$-\frac12$, respectively, which supports an interpretation as
half-instanton and half-antiinstanton.  With respect to cooling these
semi-classical objects are as quasistable as the $DD$ configurations.
Therefore, one is tempted to interprete the $D\overline{D}$ pair
as a solution of the field equation of motion, too. But so far we do not
have a clear understanding of these objects.

\medskip\noi
It is remarkable that selfdual or antiselfdual $DD$ configurations
are very rare in the deconfinement phase. $D\overline{D}$ 
mixed configurations are typical for this phase. 

In the deconfined phase the next important type of cooled configurations
are purely magnetic ones
($S_{\mathrm{magnetic}}>>S_{\mathrm{electric}}$) with
quantized action in units of $S_{\mathrm{inst}}/2$. We call them $M$
configurations.
With a lower probability also magnetic configurations with twice as
large action ($2M$ type configurations) are found.
After fixing the maximally Abelian gauge these configurations turn out
to be completely Abelian. We can identify them
as pure equally distributed magnetic fluxes related
to world-sheets of Dirac strings (`Dirac sheet') on the dual lattice.
With some rate they also emerge in the result of further
cooling of $D\overline{D}$ configurations.

Concluding we can say that the environment considered with fixed
holonomy at spatial boundaries provides an interesting pattern
of semi-classical objects characteristic for the confinement as
well as for the deconfinement phase. We have no
evidence so far, that the finite temperature gauge fields
in large volumes can be understood in terms of quantum fluctuations
around calorons with non-trivial holonomy. This question is
under consideration at present.
Anyway, we feel that the development of a semiclassical approach based on
solutions with non-trivial holonomy
might have a chance to shed more light on the mechanisms of the
deconfinement transition.

\end{document}